\documentclass[twocolumn,showpacs,prc]{revtex4}

\usepackage{amsmath}
\usepackage{amssymb}
\usepackage{amsthm}

\input{epsf}

\begin{document}

\newcommand{\Sc}{Schr\"odinger}
\newcommand{\E}{equation}
 \newcommand{\al}{\alpha }
 \newcommand{\vfi}{\varphi }

\title{\bf The phase shift effective range expansion
from supersymmetric quantum mechanics}
\author{Boris F. Samsonov$^1$
\footnote{On leave from Physics Department, Tomsk State
University, Russia, e-mail: samsonov@phys.tsu.ru} and Fl.
Stancu$^2$ \footnote{e-mail: fstancu@ulg.ac.be} } \affiliation{
$^1$Departamento de Fisica Teorica, Universidad de Valladolid,
E-47005 Valladolid, Spain \\
$^2$Universit\'e de Li\`ege, Institut de Physique B5, Sart Tilman,
B-4000 Li\`ege~1, Belgium}


\begin{abstract}
Supersymmetric or Darboux transformations are used to construct
local phase equivalent deep and shallow potentials for $\ell \neq
0$ partial waves. We associate the value of the orbital angular
momentum with the asymptotic form of the potential at infinity
which allows us to introduce adequate
long-distance transformations. The
approach is shown to be effective in getting the correct phase
shift effective range expansion. Applications are considered for
the $^1P_1$ and $^1D_2$ partial waves of the neutron-proton
scattering.
\end{abstract}


\maketitle

\section{Introduction}
In the description of the interaction between composite particles
by local potentials an ambiguity arises between different phase
equivalent or {\it isophase} potential families.
There are shallow potentials, which possess only physical bound states
and deep potentials, which possess both physical and unphysical
bound states.
The latter, called Pauli forbidden states,
simulate nonlocal aspects of the potential, or else,
the complexity of the interaction
between composite particles. The number of Pauli forbidden
states can be predicted from a microscopic description of the
interacting particles \cite{SAITO}.
The application of the supersymmetric (SUSY) quantum mechanics
\cite{WITTEN}
to the inverse scattering problem provides an
elegant and powerful algebraic way to understand the relation
between such phase equivalent potentials \cite{SUKUMAR,BS1,LEEB,SS}.
A supersymmetric transformation
can be seen as a specific Darboux transformation. In the following we shall
use Darboux and SUSY transformations as synonyms.

The above procedures cannot directly provide a correct behavior of
the phase shift at energies small relative to the potential
strength, i. e. a correct effective range expansion for $\ell >
0$. In our opinion the reason is that the role of the angular
momentum for a given central potential was not properly understood
so far.
    In the framework of SUSY quantum mechanics the
problem has been raised by Sukumar \cite{SUKUMAR}
 and  Baye and Sparenberg \cite{BS1} but not solved in principle.
However, it has been tackled
pragmatically in Ref. \cite{SPAR}. For example, in the case of the
$\ell = 4$
partial wave it led to an S-matrix containing powers of $k$
restricted to  $n \geq 10$. How one arrives at such a restriction is neither
explained nor the power 10 is justified.

The basic idea of this study is to associate
the angular momentum
with the long-distance asymptotic behavior of the potential,
irrespective of its singularity at the origin. This is in the
spirit of Ref. \cite{SWAN} where these asymptotic limits are
independent of each other.
This starting point will provide a new  possibility for getting
 a correct effective range expansion of the phase shift
which is
the following
 Taylor series expansion in the vicinity of
$k=0$ (see e. g. \cite{NEWTON})
 \begin{equation}\label{range}
 k^{2 \ell + 1} \cot \delta_{\ell}(k)=
-\frac {1}{a_{0 \ell}} +\frac {1}{2} r_{0 \ell}k^2+\ldots\,~.
\end{equation}
Here $a_{0 \ell}$ is the scattering length and $r_{0 \ell}$ the
effective range. The expression (\ref{range}) implies that for a
given $\ell$, in the series expansion of $\tan\delta_{\ell}(k)$
the coefficients of the terms containing powers of $k$ below
$2 \ell + 1$  must vanish.
In the frame of SUSY quantum mechanics
we solve this problem by introducing
adequate long-distance Darboux transformations.

The paper is organized as follows. In the subsection A of the
next section we introduce the Darboux transformation method
and briefly review the $\ell$-fixed transformations.
In subsection B we introduce 
the long-distance transformations.
Section III is devoted to results and applications to the
neutron-proton scattering. Details are worked out for the $\ell$=1
and $\ell$=2 partial waves. Conclusions are drawn in the last section.


\section{Theory}
\subsection{$\ell$-fixed Darboux transformations}

We recall that the Darboux transformation method consists in getting
solutions $\vfi$ of one Schr\"odinger equation
\begin{equation}\label{trans}
h_1\vfi =E\vfi ,\quad h_1=-\frac{d^2}{dx^2}+V_1(x)~,
\end{equation}
when solutions $ \psi$ of another equation
\begin{equation}\label{init}
h_0\psi =E\psi ,\quad h_0=-\frac{d^2}{dx^2}+V_0(x)~,
\end{equation}
are known. This is achieved by acting
on $\psi$
with a differential operator
$L$
of the form
\begin{equation}\label{L1}
\vfi =L\psi , \quad L=-d/dx +w(x)\,,
\end{equation}
where the real function $w(x)$, called {\it superpotential}, is defined as the
logarithmic derivative of a known solution of (\ref{init}) denoted by $u$
in the following. One has
\begin{equation}\label{SUPER}
w=u'(x)/u(x)\,,\quad h_0u=\al u\,,
\end{equation}
with $\al \le E_0$, where $E_0$ is the ground state energy of
$h_0$ if it has a discrete spectrum or the lower bound of the
continuous spectrum otherwise. The function $u$ is called {\it
transformation} or {\it factorization function} and $\al $ its
{\it factorization constant} or {\it factorization energy}. The
potential $V_1$ is defined in terms of the superpotential $w$ as
\begin{equation}\label{V1general}
V_1(x)=V_0(x)-2w'(x)\,.
\end{equation}
Eq. (\ref{L1}) defines a first order Darboux transformation. In
the following we shall deal with chains of $N$ successive
transformations of this type.

Let us start by first considering  $\ell$-{\it fixed}
transformations as in Ref. \cite{SS}. This means that we use a
special chain of $N=2n - \nu$ first order Darboux transformations
with $\nu\ge 0$,  generated by the following system of
transformation functions
\begin{eqnarray}\label{uv}
v_{1}(x)\,,\ldots \,, v_\nu (x)\,,
u_{\nu +1}(x)\,,  v_{\nu +1}(x)\,,\ldots \,, u_n(x)\,, v_n(x)\\
h_0u_j(x)=-a^2_ju_j(x)\,,\quad h_0v_j(x)=-b^2_jv_j(x)\,,\qquad
\end{eqnarray}
where $v_j$ are regular ($v_j(0) = 0$) and
$u_j$ irregular ($u_j(0) \neq 0)$, the latter being expressed in terms of
the Jost solutions as
\begin{equation}\label{uj}
u_j(x)=A_jf(x,-ia_j)+B_jf(x,ia_j)\,.
\end{equation}
They have arbitrary eigenvalues $-{a_j}^2$ and  $-{b_j}^2$
respectively, but always below  $E_0$.
If we are interested in the final action of the chain only, the solution
$\psi _N (x,k)$ of the transformed equation with the Hamiltonian
\begin{equation}\label{FINAL}
h_N = -d^2/dx^2+ V_N
\end{equation}
corresponding to the energy $E = k^2$ is given
by \cite{CRUM}
\begin{equation}\label{psiN}
\psi _N (x,k)=
W(u_1,\ldots ,u_N,\psi _0(x,k))\,
W^{-1}(u_1,\ldots ,u_N)
\end{equation}
where $W$ are Wronskians expressed in terms of $u_j$, denoting
symbolically any function of (\ref{uv}) and of $\psi _0(x,k)$
which is a solution of the original Schr\"odinger equation
corresponding to the same energy $E$. In the Hamiltonian
(\ref{FINAL}) the transformed potential is
\begin{equation}\label{VN}
V_N=V_0-2\frac {d^2}{dx^2}\ln W(u_1,\ldots ,u_N)\,.
\end{equation}
For $N=1$ one has $W(u_1)\equiv u_1$
and one recovers (\ref{V1general}) with $u=u_1$.
If $V_0$ is finite at the origin, $V_N$ behaves as
$\nu (\nu +1)x^{-2}$ when $x\to 0$.
Therefore the parameter $\nu$ is called the {\it singularity strength}.
The formulas (\ref{psiN}) and (\ref{VN})
result from the replacement of a chain of $N$ first
order transformations by a single $N$th order transformation,
which happens to be more efficient in practical calculations.

In Ref. \cite{SS} we obtained
that the transformed Jost function $F_N$ is related to the initial
Jost function $F_0$ by
\begin{equation}\label{MNnu}
F_N(k)=
F_0(k)\prod\limits_{j\,=\,1}^{\nu }\frac{k}{k+ib_j}
\prod\limits_{j\,=\,\nu+1 }^n\frac{k-ia_j}{k+ib_j}~.
\end{equation}
For $\nu = 0$ the first product is unity. Since a Jost function is
analytic in the upper half of the complex $k$-plane (see e. g.
\cite{FADDEEV}) all $b$'s must be positive whereas the $a$'s can
have any sign, so that every positive $a_j$ corresponds to a
discrete level $E=-a_j^2$ of $h_N$.

The corresponding phase shift $\delta^{N}_\ell (k)$ can  be written as
\begin{equation}\label{deltaN}
\delta^{N}_\ell (k)=\delta^{\,0}_\ell (k)+\Delta ^N_\ell (k) \,,
\end{equation}
where $\delta^{0}_\ell (k)$ is the initial phase shift due to the
potential $V_0$ and $\Delta^{N}_\ell(k)$ is the phase shift
produced by the chain of $N$ Darboux transformations
\begin{equation}\label{deltaNM}
\Delta ^N_\ell (k) \, =
- \,\sum _{j=\nu +1}^n\arctan (k/a_j)
- \sum\limits_{j\,=\,{1}}^n\arctan (k/b_j)~.
\end{equation}
This is consistent with the asymptotic form of
the scattering solution $\sin (kx-\frac{\pi \ell}{2}+\delta^N_\ell )$.
In the limit $k \rightarrow \infty$ one has
$\delta^{N}_\ell \rightarrow (\ell - \nu) \pi/2$,
in agreement with Ref.
\cite{SWAN} for a singular potential of parameter $\nu$. More
detailed discussion of properties of $\ell$-fixed transformations
may be found in \cite{SS}.

In the case $\ell=0$ by
expanding $\delta^{0}_\ell (k)$ and the arc tangent functions in
(\ref{deltaNM})
in power series one obtains the effective range expansion (\ref{range}).
For $\ell>0$ the situation is more subtle,
since the first term in power series of arc tangent functions is proportional to $k$.
Therefore in the usual practice based on the SUSY approach,
where $\delta^{0}_\ell (k)$
is fixed, it would be difficult to cancel the undesired powers of $k$
in order to comply with (\ref{range}). As mentioned above,
we believe that the reason is that one deals with
Darboux transformations which do not
affect
the long distance behavior of the resulting potential,
as it was pointed out in Ref. \cite{SS}.

\subsection{Long-distance Darboux transformations}
To change
the long distance behavior of a potential by SUSY
transformations we use transformation functions with {\it
zero eigenvalue}.
To cancel
undesired powers in the series expansion of arc tangent functions we
derive a proper {\it background phase shift}
as shown below.

Consider the potential $V_0(x)$ which
for $x\to \infty $
behaves as
\begin{equation}\label{V0}
V_0(x)=\frac{ {\ell} ({\ell} + 1)}{x^2}+O\left(x^{-2-\gamma }\right)\,,\quad
\gamma >0,\quad \ell \ge 0  \,.
\end{equation}
As it is known \cite{FADDEEV}
the Schr\"odinger equation containing a potential satisfying
(\ref{V0})
has zero eigenvalue solutions with the following asymptotic behavior at
$x\to \infty $
\begin{eqnarray}\label{as1}
v(x)&=&Cx^{\,\ell\,+1}\left[ 1+O\left(x^{-\,\gamma }\right) \right] \\
\label{as2}
u(x)&=&\frac{D}{x^{\,\ell}}\left[\, 1+O\left(x^{-\,\gamma }\right) \right]\,.
\end{eqnarray}
The functions (\ref{as1}) are regular at the origin but singular at infinity
and the functions (\ref{as2}) are
just the other way round.
When these functions are taken as
transformation functions
the change in the
potential
for sufficiently large $x$ has one of the following forms
\begin{eqnarray}\label{v01}
\Delta V(x)&=&
-2[\ln v(x)]''=\frac {2({\ell}+1)}{x^2}+o(x^{-2-\gamma }) \\
\label{v02}
\Delta V(x)&=&
-2[\ln u(x)]''=-\frac {2 {\ell}}{x^2}+o(x^{-2-\gamma })
\end{eqnarray}
where $\Delta V(x) = V_N - V_0$ with $N$ = 1 in (\ref{VN}). It is
clear from here that the function (\ref{as1}) increases the value
of $\ell$ by one unit and the function (\ref{as2}) decreases it by
one unit. Moreover a linear combination of (\ref{as1}) and
(\ref{as2}) is a function of type (\ref{as1}). They form a
one-parameter family, while the function (\ref{as2}) is uniquely
defined (up to an inessential constant factor). In this family
there is only one function regular at the origin. This function,
used as transformation function in the Darboux algorithm, changes
both $\nu $ and $\ell$ but all the other members of the singular
family change $\ell$ without affecting $\nu $. In the following we
shall use the singular functions of the one-parameter family
defined above to derive phase shifts leading to a correct
effective range expansion. We shall therefore show that the
parameters appearing in the linear combination of (\ref{as1}) and
(\ref{as2}) can be chosen such as the resulting phase shift
provides the general effective range expansion (\ref{range}).
Hence, starting with a given $V_0$ with $\ell = 0$ we first
perform a number $N=\ell$ of 
transformations which
give the correct long distance behavior of the potential
$V_{\ell}$ and introduce $\ell$ parameters in the phase shift.
Next, an $\ell$-fixed chain is performed, producing the final
phase shift (\ref{deltaN}) for which the potential $V_{\ell}$
plays the role of the initial potential. The latter transformation
does not affect the asymptotic form of the potential $V_\ell$ at
large $x$. Hence, the resulting potential $V_N$ has an asymptotic
behavior corresponding to the $\ell$th partial wave. The addition
of $\ell$ zero-energy eigenfunctions to the $N=2n-\nu$ first order
transformation functions used in the $\ell$-fixed chain increases
$N$ by $\ell$ units. This means that in the $N$th order
transformation to be used below the total number of transformation
functions is
\begin{equation}\label{NTOT}
N = 2 n -\nu + \ell~.
\end{equation}
If we start with the zero initial potential, $V_0\equiv 0$,
the formula (\ref{deltaN})   for the phase shift
has to be modified as follows:
\begin{equation}\label{deltaNnew}
\delta^{N}_\ell (k)=\delta^{\ell}_\ell (k)+ \Delta ^{N-\ell}_\ell (k) \,.
\end{equation}
 Here $\delta^{\ell}_\ell (k)$ is
 produced by  
 the long-distance transformations which give rise to
 an intermediate potential $V_\ell$.
  In the following $\delta^{\ell}_\ell (k)$ will play the role
 of a {\it background phase shift}.
 The additional phase shift $\Delta ^{N-\ell}_\ell (k)$
 corresponding to the $\ell$-fixed subchain of $N-\ell =2n-\nu $
 transformations has the same form as (\ref{deltaNM}).
 We shall illustrate this procedure by applications given in the following
 section.

\section{Applications}

\subsection{\it {\bf The case $\ell =1$}}
We start with the potential $V_0=0$ in Eq. (\ref{init}).
Let us take $u_{0,1}=x+x_0$ as the transformation function
with zero eigenvalue,
where $x_0 \geq 0$ is a free parameter.
Then the first order transformation operator (\ref{L1}) takes the form
\begin{equation}\label{L01}
L_{1}=-\frac d{dx}+\frac 1{x+x_0}~.
\end{equation}
The
transformed potential is
\begin{equation}\label{Vl1}
V_{1}=\frac 2{(x+x_0)^2}\,,\quad x_0\ge 0
\end{equation}
and its Jost solution
$f_1(x,k)$ may be found by applying the operator (\ref{L01}) on
the Jost solution $f_0(x,k)=\exp (ikx)$ of the free particle equation.
After dividing by the factor $ik$ one finds
\begin{equation}\label{Jost10}
f_{1}(x,k)
=\left(1-\frac{1}{ik(x+x_0)}\right)
\exp (ikx)~.
\end{equation}
For $x_0 \neq 0$ the potential (\ref{Vl1}) has $\nu =\nu_{1}=0$
and its Jost function $F_{1}(k)$ coincides with $f_{1}(0,k)$.
If
one now applies the operator (\ref{L01}) on an oscillating solution of
the free particle equation $\sin(kx + \delta_{1}^1)$ one obtains
\begin{equation}\label{tildevfi}
\varphi_{1} (x,k)= -k\cos (kx+\delta_{1}^1)+
\displaystyle {\frac{\sin (kx+\delta_{1}^1)}{x+x_0}}\,.
\end{equation}
This solution is regular at the origin provided
\begin{equation}\label{d01}
\delta_{1}^1=\arctan kx_0 \,,
\end{equation}
and has the asymptotic behavior $\sim\sin (kx+\delta_{1}^1-
{\textstyle\frac{\pi}{2}})$ at $x\to \infty$, it describes the
$\ell = 1$ scattering state of the potential ({\ref{Vl1}). Thus we
have switched from the partial wave $\ell = 0$ of $V_0 = 0$ to
$\ell = 1$ of the potential (\ref{Vl1}). Now we can perform
$\ell$-fixed transformations. The free parameter $x_0$ will be
chosen so that the final phase shift (\ref{deltaNnew}) will have the
correct effective range expansion.
Replacing $\delta^{\ell}_\ell$ by the result
(\ref{d01}) and expanding all arc tangent functions in power
series one can see that the coefficient of the term proportional
to $k$ in $\tan \delta^N_1$ vanishes for
\begin{equation}\label{x0}
x_0=\sum\limits_{j\,=\nu+1}^n a_j^{-1} + \sum\limits_{j\,=1}^n b_j^{-1}\,.
\end{equation}
Both the regular and irregular solutions
corresponding to  the potential (\ref{Vl1})
can be found with the help of the Jost solution (\ref{Jost10}).
But in the spirit of Ref. \cite{CRUM}, we can
avoid this step, thus considerably reducing the amount of numerical work.
This means that in
the formulas (\ref{psiN}) and (\ref{VN})
we can directly use
appropriate solutions of the free particle equation
which are simple linear combinations of exponentials
$$
\psi_j(x,k) = A_j  \exp(ik_jx) + B_j \exp(-ik_jx)\,, \quad
\mbox{Re}\,k_j = 0\,,
$$
where we have to find the correct ratio $B_j/A_j$.
Since the regular solutions of the potential (\ref{Vl1}),
consisting of
functions $v_j(x)=L_1\psi_j(x,-ib_j)$
satisfy the condition $v_j(0)=0$, which fixes the
ratio $B_j/A_j$, we have free particle solutions of the form
\begin{equation}\label{vax0}
\psi_j(x,-ib_j) =(b_jx_0+1)\exp (b_jx)+
(b_jx_0-1)\exp (-b_jx)\,.
\end{equation}
The irregular solutions for the same potential, defined as
$u_j=L_1\psi_j(x,-ia_j)$, should be obtained from the functions
\begin{equation}\label{virr}
\psi_j(x,-ia_j)=A_j\exp (a_jx)+B_j\exp (-a_jx) \,,
\end{equation}
which for
 $B_j/A_j\ne (a_jx_0-1)/(a_jx_0+1)$, $A_j\ne 0$ and
 $a_j>0$  have an increasing asymptotic behavior but if
$A_j=0$ they  decrease asymptotically.
 To stress the difference between solutions
 with different asymptotic behavior
  we choose in the latter case $a_j<0$ and $A_j\ne 0$ but $B_j=0$
 (for more details see \cite{SS}).
 Here $a_j$, $b_j$ and $x_0$ are parameters of the model
 to be found below.
  Then
in the $N$th order transformation
we use the functions
$u_{0,1}=x+x_0$,
 (\ref{vax0})
and (\ref{virr}) to calculate  (\ref{VN}).
 Note nevertheless that in the
resulting phase shift, given by (\ref{deltaNnew}),
$\delta^\ell_\ell $ has to be replaced by
$\delta_1^1$ of (\ref{d01}) since the initial potential for the
subchain of $\ell$-fixed transformations is now $V_1$ of
(\ref{Vl1}).
For the $N$th order transformation the potential $V_1$ and
the phase shift $\delta_1^1$ play an auxiliary role.


\begin{table}
\begin{ruledtabular}
\caption{ The $^1P_1$ phase shift. The theoretical value is
calculated from Eq. (\ref{deltaN}) with N = 7 transformation
functions, $E_{lab} = \frac{m_n + m_p}{m_n} E_{cm} $,
\protect{$E_{cm} = \frac{{\hbar}^2 k^2}{2 \mu}$} where $\mu$ is
the reduced mass, $m_p = 938.27$ MeV and $m_n = 939.36$ MeV. The
experimental values are from Ref. \protect\cite{STOKS}
\label{table1}}
\begin{tabular}{ccc}
 \hline
$E_{lab}(MeV)$ & $\delta^{exp}_1(deg)$ & $\delta^7_1(deg)$ \\
 \hline
14        & -4.1944             & -4.27887            \\
42        & -9.01021            & -8.91287          \\
70        & -11.99126           & -11.9844           \\
98        & -14.42546           & -14.5092           \\
126       & -16.60093           & -16.7024            \\
154       & -18.59407           & -18.6533            \\
182       & -20.42528           & -20.4137            \\
210       & -22.09942           & -22.0187            \\
238       & -23.61771           & -23.4936            \\
266       & -24.9813            & -24.8577            \\
294       & -26.19215           & -26.1263            \\
322       & -27.25323           & -27.3113             \\
350       & -28.16843           & -28.4228             \\
\hline
\end{tabular}
\end{ruledtabular}
\end{table}

\begin{table}
\begin{ruledtabular}
\caption{Theoretical values of
the scattering length and the effective range. \label{table2} }
\begin{tabular}{ccccc}
 \hline
$a_{01}$ & $r_{01}$  & $a_{02}$ & $r_{02}$ & Reference\\
\hline
3.143 &  - 6.302 & - 2.224 & 21.976 & present work \\
3.023 &  - 6.895 & & &  \protect\cite{RIJKEN} \\
2.736 &  - 6.449 & - 1.377 & 15.027 & Reid93 (\protect\cite{MART} \\
2.4 $\pm$ 1.3 & -12.6 $\pm$ 2.2 & & & (\protect\cite{MATHELITSCH})\\
\hline
\end{tabular}
\end{ruledtabular}
\end{table}

As an application we look for neutron-proton ($np$) potentials
which reproduce the 'pruned' phase shift of
\cite{STOKS} for the ${}^1P_1$ partial wave.
This phase shift together with the theoretical values obtained
from the expression (\ref{deltaN}) and
denoted by $\delta^{7}_1$ are exhibited in Table \ref{table1}.
The six fitted $S$-matrix poles are
\begin{equation}\label{1ppolesx0}
\begin{array}{l}
a_1=-0.7290,\ a_2=-0.7295,\ a_3=1.0368\\
b_1=\hphantom{-}0.4403 ,\ \  b_2=\hphantom{-}0.4408,\  b_3=3.3818
\end{array}
\end{equation}
in fm$^{-1}$ units. The superscript 7 carried by the phase shift
is consistent with the formulas (\ref{NTOT}) and
 (\ref{deltaNnew})
 and it implies that we
used a $7$th order transformation
according to   (\ref{psiN}) and (\ref{VN}).
Then from Eq. (\ref{x0}) one gets $x_0 =
3.0578$ fm. Now we can expand all seven arc tangent functions
appearing in (\ref{deltaN}) in power series. This leads to a
correct effective range expansion given by
\begin{equation}\label{ER01}
k^3\cot \delta^{7}_1(k)= - 0.3182 - 3.1511 k^2  +\ldots \,
\end{equation}
from which one can extract the scattering length $a_{01}$ and the effective
range $r_{01}$ defined according to (\ref{range}).
 In Table \ref{table2} these values are compared with
another theoretical model \cite{RIJKEN}. They are surprisingly
close to each other. The phenomenological Reid93 potential
\cite{STOKS} also gives similar values \cite{MART}. Moreover the
scattering length is located in the interval deduced from a
partial wave analysis \cite{MATHELITSCH}.

In order to construct potentials giving rise to the phase shift
$\delta^{7}_1$, the poles (\ref{1ppolesx0}) have to be associated
with transformation functions defined  by (\ref{vax0}) and
(\ref{virr}). The poles $b_j$ correspond to regular solutions
of $V_1(x)$
 resulting from (\ref{vax0}).
 The poles $a_1$ and $a_2$, which
are negative, correspond to decreasing functions
 of the form
(\ref{virr}) with $B_{j}=0$ ($j$=1,2). It remains the pole $a_3$
which is positive. If we take $B_3/A_3 \ne (a_3x_0-1)/(a_3x_0+1)$
in Eq. (\ref{virr}),
we obtain from (\ref{VN})
a one-parameter
($B_3/A_3$) family of one-level isophase deep potentials with the
discrete level at $E=-a_3^2$. But if we choose $B_3/A_3 =
(a_3x_0-1)/(a_3x_0+1)$ the initially irregular
function
 moves into the regular
family,
\begin{figure}
\epsfysize=5cm \epsffile{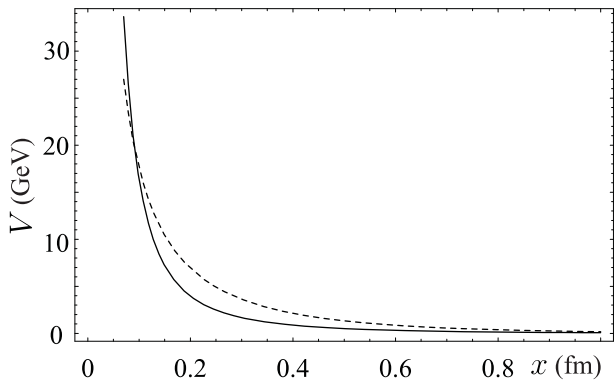} \caption{\small ${}^1P_1$
$np$ shallow potentials. $V_7$ (solid line) together with the
Reid68 potential \protect\cite{REID68} (dashed line).}
\label{fig1}
\end{figure}
\begin{figure}
\epsfysize=5cm \epsffile{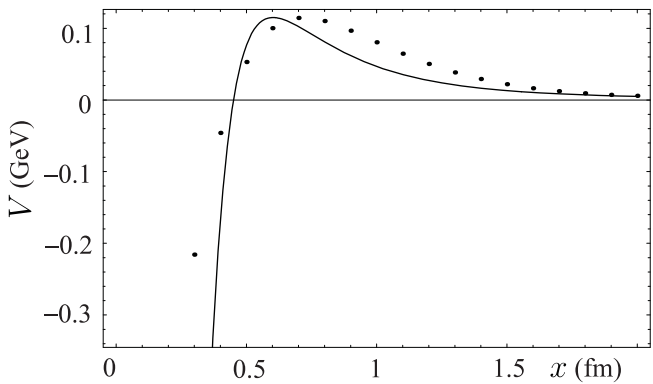} \caption{\small ${}^1P_1$
$np$ deep potentials. The full line corresponds to the
potential which gives the correct
effective range expansion (\ref{ER01}) and the dots
represent the Reid93 potential \protect\cite{STOKS}. }
\label{fig2}
\end{figure}
the level disappears and one gets the uniquely defined shallow
representative of the family of isophase potentials, denoted by
$V_7$.
Fig. \ref{fig1} shows this potential from which the
centrifugal term has been subtracted. The potential $V_7$,  is
quite close to the Reid68 potential \cite{REID68}, represented in
the same figure. Figure \ref{fig2} shows one of the deep isophase
potentials corresponding to $B_3/A_3 = 0.351$. This constant has
been adjusted to get a potential as close as possible to Reid93
\cite{STOKS} in the interval $ 0.4$ fm $\leq x \leq 2$ fm. This
deep potential possesses a Pauli forbidden state of energy $E = -
a_3^2  = - 44.46$ MeV. Contrary to the Reid93 potential which is
deep but finite, our potential behaves at origin as $-2/x^2$.  If
necessary, it can be regularized as for example in Ref.
\cite{SPAR}.

It would be interesting to analyze the asymptotic behavior of our
calculated potential to see if it is compatible with modern
phenomenological potentials constructed in the spirit of meson
theory, i.e. which include one-pion-exchange (OPE) contributions.
This is precisely the case of the Reid93 potential, which includes
OPE with neutral and charged pions (for details see Ref.
\cite{STOKS}). The comparison given in Figure \ref{fig3} shows
that the shallow potential $V_7$ and
its deep partner are practically
identical in the asymptotic region
and that they are both extremely close to the
asymptotic form of the updated Reid93 potential \cite{MART}. Thus
our potentials are in the excellent agreement with expectations from Yukawa's
OPE theory.  We remind that we subtracted the centrifugal barrier
from our potentials in order to make this comparison feasible. In
solving the Schr\"odinger equation this should be added back to the
nucleon-nucleon interaction in each case. The fact that the shallow
and deep potentials are asymptotically the same is physically
correct. As mentioned in the introduction, deep potentials
reflect the compositeness (quark structure) of the interacting particles
in the overlap region but, once the
particles are well separated, they could be treated
as point particles as in  OPE theories, which means that beyond
some distance the deep and shallow potentials should coincide.
\begin{figure}
\epsfysize=5cm \epsffile{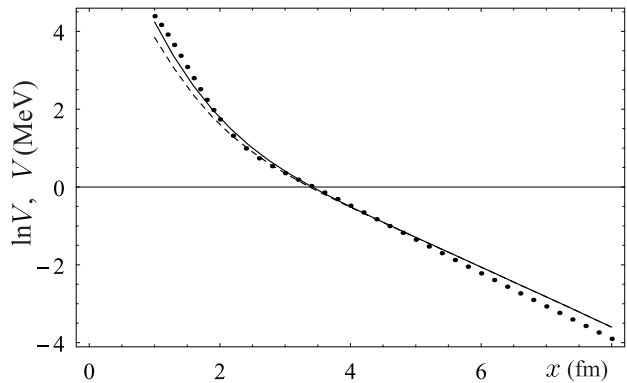} \caption{Asymptotic behavior
in natural logarithmic scale of the value of the \small ${}^1P_1$
$np$ potentials. The full line represents our shallow potential of
Fig. 1, the dashed line our deep potential of Fig. 2 and the
dotted line the updated \protect\cite{MART} phenomenological
Reid93 \protect\cite{STOKS} potential, including one-pion
exchange. } \label{fig3}
\end{figure}

\subsection{\it \bf{The case $\ell=2$.}} Now we have to apply two
subsequent
transformations associated with functions corresponding to $E=0$
and then, as above, a subchain of $\ell$-fixed transformations.
After the first transformation with
the function $u_{0,1}=x+x_0$, the potential
$V_{1}$ of (\ref{Vl1}) has $u=1/(x+x_0)$ and
$\tilde u=(x+x_0)^2$
as linearly independent solutions at $E=0$.
Their linear combination
$u_{0,2}=cu+\tilde u$,
which is the transformation function for the second transformation
step defined by the operator
\begin{equation}\label{L2}
L_2=-d/dx +u'_{0,2}(x)/u_{0,2}(x)
\end{equation}
contains two free parameters $x_0$ and $c$. These can be
chosen such as the series expansion for $\tan \delta_2(k)$ starts
at $k^5$.
The intermediate (or background) potential
\begin{equation}\label{V20}
V_{2}=-2[\ln~ u_{0,2}(x) u_{0,1}(x)]\,''
=\frac{6(x+x_0)[(x+x_0)^3-2c]}{[c+(x+x_0)^3]^2}
\end{equation}
obtained from the $L=L_2 L_1$ Darboux transformation
(for more details see \cite{SS})
plays now
the role of the initial potential for an $\ell$-fixed subchain of
transformations.
The background phase shift (modulo $\pi $)
corresponding to $V_{2}$ is
\begin{equation}\label{delta02}
\delta_{2}^2 =
\arctan \frac{3kx_0^2}{3x_0-k^2(x_0^3+c)}~.
\end{equation}
Note that the function $\varphi_2 (x,k) = L_{2}L_{1}\sin
(kx+\delta _{2}^2)$
is regular at the origin and describes an
unnormalized scattering state for $\ell = 2$.
In the formula
(\ref{deltaNnew})
we have to identify  $\delta^\ell_\ell (k)$ with
$\delta _{2}^2(k)$
of (\ref{delta02}). After expanding in power series all
arc tangent functions we find that the coefficient of the term
linear in $k$ vanishes for $x_0$ given by (\ref{x0}) and
\begin{equation}\label{k3}
c = {\textstyle\frac 13}\,
[\,\sum\limits_{j\,=1}^n(a_j^{-3}+b_j^{-3})\,]\,
\end{equation}
ensures the cancellation of the coefficient of $k^3$ in the 
series expansion of $\tan \delta^N_2$.
Now,
 to find solutions  of the free particle equations,
 giving rise to the regular family of the potential
 (\ref{V20}), to be used in (\ref{VN}), we need
 eigenfunctions of
$h_0=-d^2/dx^2$
satisfying the condition
$L_{2}L_{1} \psi_j(x,-ib_j)=0$ at $x = 0$.
They
are given by the following linear combination of
exponentials
\begin{eqnarray}\label{regulv}
\psi_j(x,-ib_j)&=&[\,3x_0+3b_j x_0^2+b^{2}_j (c+x_0^3)\,]\exp (b_j x) \nonumber\\
&-&[\,3x_0-3b_j x_0^2+b^{2}_j(c+x_0^3)\,]\exp(-b_j x)\,. \nonumber \\
\end{eqnarray}
The irregular family still results from (\ref{virr})
subject to the condition that the ratio $B_j/A_j$ is different from that
presented in (\ref{regulv}).

With a fit of a similar quality to that performed for $^1P_1$  we could reproduce
the $^1D_2$ partial wave phase shift of \cite{STOKS}
with the following four poles of
the $S$-matrix
\begin{equation}\label{1dx0poles}
\begin{array}{l}
a_1=- 0.2047,\ \ a_2=-1.9800\\
b_1=\hphantom{-}1.2305,\ \ \ b_2=\hphantom{-}4.9631
\end{array}
\end{equation}
in fm$^{-1}$ units. From Eqs. (\ref{x0}) and (\ref{k3}) we get $x_0=- 4.375$ fm,
$c=116.08$ fm$^3$. This leads to the following
effective range expansion
\begin{equation}\label{DEL2}
k^5 \cot \delta^6_2 (k)=0.4496 + 10.9878 k^2+\ldots \,.
\end{equation}
where the superscript $N=6$ represents 4 transformation functions associated
with the poles (\ref{1dx0poles}) plus two zero-eigenvalue functions
$u_{0,1}$ and $u_{0,2}$ defined above. This is consistent with the
formula (\ref{NTOT}) with $n=2$, $\nu=0$ and $\ell=2$. From the
expansion (\ref{DEL2}) one can extract the scattering length
$a_{02}$ and the effective range $r_{02}$ defined in Eq. (1).
These values are shown in Table \ref{table2}. They are comparable
to those obtained for the potential Reid93 \cite{MART}.

\section{Conclusions}

By working out these two particular cases we have shown that
a new insight emerges into the role of the angular momentum
of a central potential. If associated
with the long-distance behavior of the potential, it
allows us to introduce
transformations
 that bring $\ell$ free
parameters in the background phase shift (Eqs (\ref{d01}) and
(\ref{delta02})).
When a correct effective range expansion is required,
each extra unit of angular momentum
imposes a new constraint on the whole system of parameters
of the model such as the number of constraints coincides with the
number of parameters in the background phase shift.
 For the particular cases of $\ell =1$ and $\ell =2$ explicit
 solutions of the constraint equations are given.
Thus, the extension of the method to $\ell > 2$ is straightforward.

Before ending we should mention that the generalized Levinson theorem
\cite{SWAN} is always satisfied in our approach.

\vspace{0.5cm} We are grateful to Mart Rentemeester for useful
correspondence. B.F.S. acknowledges support from the Spanish
Ministerio de Education, Cultura y Deporte Grant SAB2000-0240 and
the Spanish MCYT and European FEDER grant BFM2002-03773. He is
also grateful for hospitality at the Fundamental Theoretical
Physics Laboratory of the University of Liege.



\begin{thebibliography}{99}

\bibitem{SAITO} S. Saito, Prog. Theor. Phys. {\bf 41}, 705 (1969).

\bibitem{WITTEN} E. Witten, Nucl. Phys. {\bf B188}, 51 (1981).

\bibitem{SUKUMAR} C. V. Sukumar, J. Phys. A {\bf 18}, 2917 (1985);
{\bf 18}, 2937 (1985).

\bibitem{BS1} J. -M. Sparenberg and D. Baye, Phys. Rev. C {\bf 55}, 2175 (1997);
 J. -M. Sparenberg, D. Baye and H. Leeb, Phys. Rev. C {\bf 61},
 024605 (2000).

\bibitem{LEEB}
H. Leeb and D. Leidinger, Few-Body Syst. Suppl. {\bf 6}, 117 (1992);
R. M. Adam, H. Fiedeldey, S. A. Sofianos and H. Leeb,
Nucl. Phys. {\bf A559}, 157 (1993).

\bibitem{SS} B. F. Samsonov and Fl. Stancu, Phys. Rev. C {\bf 66},
034001 (2002).

\bibitem{SPAR} J. -M. Sparenberg, Phys. Rev. Lett. {\bf 85}, 2661 (2000).

\bibitem{SWAN} P. Swan, Nucl. Phys. {\bf 46}, 669 {1963}.

\bibitem{NEWTON} R. G. Newton {\it Scattering theory of waves and particles},
McGraw-Hill, New York, 1966.

\bibitem{CRUM} M. N. Crum, Quarterly J. Math {\bf 6}, 121 (1955).

\bibitem{FADDEEV} L.D. Faddeev, Uspehi Mat. Nauk, {\bf 14}, No
4 (88), 57 (1959);
B.M. Levitan, {\it Inverse Sturm-Liouville problems}, (Nauka,
Moscow, 1984).

\bibitem{REID68} R. V. Reid, Jr., Ann. Phys. (N.Y.) {\bf 50}, 411 (1968).

\bibitem{STOKS} V. G. J. Stoks, R. A. M. Klomp, C. P. F. Terheggen and
J. J. de Swart, Phys. Rev. {\bf C49}, 2950 (1994); for an updated
data base and  Reid93 potential
see http://nn-online.sci.kun.nl.

\bibitem{RIJKEN} M. N. Nagels, T. A. Rijken and J. J. de Swart,
in Lecture Notes in Physics, vol. 82, Few-Body Systems and Nuclear Forces
I, eds. H. Zingl, M. Haftel and H. Zankel (Springer, Berlin, 1878), p.17

\bibitem{MART} M. Rentmeester, private communication

\bibitem{MATHELITSCH} L. Mathelitsch and B. J. VerWest, Phys. Rev.
{\bf C29}, 739 (1984).



\end{thebibliography}
\end{document}